\documentclass[twocolumn,secnumarabic,amssymb, nobibnotes, aps, prd]{revtex4}
\usepackage{graphicx}

\begin{document}
\title{Could quantum mechanics describe completely and consistently all superconducting and other quantum phenomena?}
\author{A.V. Nikulov}
\email[]{e-mail: nikulov@iptm.ru}
\affiliation{Institute of Microelectronics Technology and High Purity Materials, Russian Academy of Sciences, 142432 Chernogolovka, Moscow District, RUSSIA.} 
\begin{abstract}Canonical description of quantization effects observed at measurements on superconducting structures seems one of the most triumphant achievements of quantum mechanics. But impartial consideration of this description uncovers incompleteness and inconsistency of this description. Contradictions in the description of other quantum phenomena are revealed also.    
\end{abstract}

\maketitle 

\narrowtext

\section{Introduction}
Quantum mechanics is the most successful theory of physics. The progress of physics of the last century is undeniably connected with quantum mechanics. But John Bell said that "{\it This progress is made in spite of the fundamental obscurity in quantum mechanics. Our theorists stride through that obscurity unimpeded... }" (see p. 170 in \cite{Bell2004}) because "{\it they are likely to insist that ordinary quantum mechanics is just fine 'for all practical purposes'}" \cite{Bell1989}. Bell \cite{Bell1989} as well other critics of quantum mechanics agreed with them about that ordinary quantum mechanics describes successfully all or almost all quantum phenomena. The recent publication \cite{PhysicaC2015} calls this general confidence in question. In this paper reader's attention is drawn on other examples of incompleteness and inconsistency of the universally recognized description of superconducting and other quantum phenomena.

\section{What is the 'force' propelling the mobile charge carriers to move in direction opposite to the electromagnetic force? }
\label{sec:2}
Bohr postulated as far back as 1913 that angular momentum $m _{p} = rp$ of electron in atom should have discrete values $m _{p} = n\hbar$. This postulate was extended on other cases, for example of the case of electron (or other particles) moving free along an one-dimensional ring. The quantization of angular momentum in this state described with the wave function $\Psi = |\Psi |e^{i\varphi }$ 
$$m _{p} =  \oint _{l} dl sr\Psi ^{*} (-i\hbar \nabla ) \Psi = sr|\Psi |^{2}\hbar \oint _{l} dl \nabla \varphi =  \hbar n \eqno{(1)}$$
may be deduced from the requirement $\oint _{l} dl \nabla \varphi = n2\pi $ that the complex wave function must be single-valued at any point of the ring $\Psi = |\Psi |e^{i\varphi } =  |\Psi |e^{i(\varphi + n2\pi )} $. $|\Psi |^{2} = 1/s2\pi r$ is the probability density (according to Born's interpretation of the wave function) in a homogeneous one-dimensional ring with a section $s$ and a radius $r$.  According to the prevalent definition, the operator of canonical momentum is the same $\hat{P} = -i\hbar \nabla $ with and without magnetic field \cite{LandauL}.  The operator of the velocity of a particle with a charge $q$ is $\hat{v} = (\hat{P} - qA)/m$ and the Hamiltonian is
$$\hat{H} = \frac{1}{2m}(-i\hbar \nabla  - qA)^{2} +U   \eqno{(2)}$$
in the presence of a magnetic vector potential $A$ \cite{LandauL}. Therefore, according to the prevalent definition \cite{LandauL}, the current $I_{p} = sq|\Psi |^{2}v$ of a charge particle  
$$I_{p} = \frac{sq}{m2\pi r}\oint_{l}dl \Psi ^{*}(-i\hbar \nabla  - qA)\Psi  = \frac{n\Phi_{0} - \Phi}{L_{k}}  \eqno{(3)}$$
can not be equal zero when the magnetic flux inside the ring $\Phi = \oint_{l}dl A$ is not divisible $\Phi \neq n\Phi _{0}$ by the flux quantum $\Phi _{0} = 2\pi \hbar /q$ and only the kinetic energy $E_{k}$ in 
$$\int _{V}dV\Psi ^{*}H\Psi =  \frac{(n\Phi_{0} - \Phi )^{2}}{2L_{k}} + \int _{V}dV\Psi ^{*}U\Psi \eqno{(4)}$$
depends on the magnetic flux $\Phi $. $L_{k} = ml/sq^{2}|\Psi |^{2}$ is the kinetic inductance of the ring with a length $l = 2\pi r$, a section $s$ and a density $|\Psi |^{2}$ of particles with a charge $q$. 

The prevalent definition of the momentum operator and the Hamiltonian (2) are used for the description of quantization effects observed in superconductors \cite{QuTh1961}. Feynman noted fairly  that {\it "in a situation in which $\Psi $ is the wave function for each of an enormous number of particles which are all in the same state, $|\Psi |^{2}$ can be interpreted as the density of particles"} \cite{FeynmanL}. Superconductivity is just such situation. All superconducting  pairs are all in the same state in a superconductor. Just therefore this macroscopic quantum phenomenon can be observed \cite{FPP2008}. Schrodinger's interpretation rather than Born's interpretation of the wave function should be used for the description of quantization effects observed in superconductors. The value $|\Psi |^{2}$ is the real density of superconducting  pairs $n _{s}$ according to this interpretation. The relations (3) and (4) are used for the description of quantum periodicity in different parameters observed at measurements of superconducting rings (or loop) with small section $s \ll \lambda _{L}^{2}$ \cite{Letter07,toKulik2010,PCScien07,JETP07J,PCJETP07,PerMob2001,Letter2003,PLA2012}. $\lambda _{L} = (m/\mu _{0}q^{2}n_{s})^{0.5} = \lambda _{L}(0)(1 - T/T_{c})^{-1/2}$ is the London penetration depth, $\lambda _{L}(0) \approx 50 \ nm = 5 \ 10^{-8} \ m$ for most superconductors \cite{Tinkham}. The kinetic inductance $L _{k} \approx (\lambda _{L}^{2}/s) \mu _{0}l$ exceeds the magnetic inductance $L_{f}  \approx \mu _{0}l$ in this case of weak screening. One can always neglect the magnetic flux $\Delta \Phi _{I} = L_{f} I _{p}$ induced with the current $I _{p}$ for a sufficiently thin superconductor with $s \ll \lambda _{L}^{2}$ \cite{Tinkham}. The magnetic flux $\Phi = \Phi_{ext} + L_{f} I _{p}$ equals approximately  $\Phi \approx \Phi_{ext}$ the one $\Phi_{ext} = BS $ of externally produced magnetic field $B$ at $L_{f} \ll  L_{k}$. The quantization (1) may be used also for the description of magnetic flux quantization \cite{FQ1961} and Meissner's effect \cite{Meissner1933} observed in the case of strong screening, when superconductor size $w$ is large $w \gg \lambda _{L}$ \cite{FPP2008}. 

According to the universally recognized explanation \cite{Tinkham} quantum periodicity in the transition temperature \cite{LP1962,LP2001}, the ring resistance \cite{Letter07,toKulik2010}, its magnetic susceptibility \cite{PCScien07}, the critical current  \cite{JETP07J} and the dc voltage measured on segments of asymmetric rings \cite{Letter07,toKulik2010,PerMob2001,Letter2003,PLA2012} are observed due to the change of the quantum number $n$ with magnetic flux at $\Phi = (n'+0.5)\Phi_{0}$. The quantum number $n$ changes because the energy (4) is minimal and the superconducting state has maximal probability $P _{n}$ at $n = n'$ when $\Phi < (n'+0.5)\Phi_{0}$ and at $n = n'+1$ when $\Phi > (n'+0.5)\Phi_{0}$ \cite{Tinkham}. The two state $n = n'$ and $n = n'+1$ have the same value of the kinetic energy in (4) $E _{k}= (n\Phi_{0} - \Phi )^{2}/2L_{k} = \Phi_{0}^{2}/8L_{k}$ at $\Phi = (n'+0.5)\Phi_{0}$. The fractional depression of the transition temperature depends of the kinetic energy (4) $\Delta T _{c}/T _{c} \propto  -E _{k} \propto -(n\Phi_{0} - \Phi)^{2}$ \cite{Tinkham}. Therefore the maximums of the $T _{c}(\Phi)$ oscillations are observed at  $\Phi = n'\Phi_{0}$ and the minimums at $\Phi = (n'+0.5)\Phi_{0}$ \cite{LP2001}. The oscillations $\Delta R(\Phi) \propto (n\Phi_{0} - \Phi)^{2}$ \cite{Letter07,toKulik2010} measured in the fluctuation region near the transition temperature, where the resistance changes from $R = 0$ at $T < T _{c}$ to $R = R _{n}$ at $T > T _{c}$, are considered as a consequence of the $T _{c}(\Phi)$ oscillations \cite{LP1962}. The magnetic susceptibility measured in the fluctuation region equals zero at $\Phi = n'\Phi_{0}$ and $\Phi = (n'+0.5)\Phi_{0}$  \cite{PCScien07} because it is proportional to the persistent current average in time $\Delta \Phi _{Ip} = L_{f}\overline{I_{p}}$: $\overline{I_{p}} \approx (n'\Phi_{0} - \Phi)/L _{k} = 0$ at $\Phi = n'\Phi_{0}$ and $\overline{I_{p}} \approx P _{n'} (n'\Phi_{0} - \Phi)/L _{k} + P _{n'+1}[(n'+1)\Phi_{0} - \Phi]/L _{k}  = 0$ because $P _{n'} = P _{n'+1}$ at $\Phi = (n'+0.5)\Phi_{0}$. The persistent current (3) corresponding to the minimal energy (4) is diamagnetic at $n'\Phi_{0} < \Phi < (n'+0.5)\Phi_{0}$  and paramagnetic at $(n'+0.5) \Phi_{0} < \Phi < (n'+1)\Phi_{0}$. Magnetic field dependence of the critical current $I_{c}(\Phi )$ of a symmetrical ring has the maximums at  $\Phi = n'\Phi_{0}$ and the minimums at $\Phi = (n'+0.5)\Phi_{0}$, see Fig.2 in \cite{JETP07J}, because the persistent current increases the total current in one of the ring halves, see Fig.1 in \cite{PCJETP07}, and therefore $I_{c} = I_{c0} - 2|I_{p}| = I_{c0} - 2|n\Phi_{0} - \Phi |/L _{k}$. The dc voltage oscillations $V_{dc}(\Phi ) $ are observed due to the rectification of the ac current \cite{Letter2003,PCJETP07} or a noise \cite{Letter07,toKulik2010,PerMob2001,PLA2012} observed at measurements of asymmetric rings. The dc voltage changes its sign at $\Phi = n'\Phi_{0}$ and $\Phi = (n'+0.5)\Phi_{0}$ as well as the average value of the persistent current $\overline{I_{p}}$. Thus, Bohr's quantization (1) and the influence of the magnetic vector potential $A$ on the phase $\nabla \varphi $ of the wave function (call sometimes as Aharonov-Bohm effect \cite{ABNanoSt2009}) seem to describe successfully numerous quantum phenomena observed in superconducting rings and also in normal metal mesoscopic rings \cite{PCScien09,PCPRL09}. 

\subsection{Transition between continuous and discrete spectrum of permitted states.}
\label{sec:2.1}
Some authors \cite{ArtAt2010,ArtAt2011,ArtAt2012} consider superconducting loop as an artificial atom because its spectrum of permitted states is discrete due to Bohr's quantization. These artificial atoms provide additional experimental opportunities for studies of quantization phenomena. We can, for example, to observe a transition from continuous to discrete spectrum of permitted states of superconducting loop, which can not be observed in the case of atom. This transition is observed when whole ring \cite{PRB2014C} or its segment \cite{JLTP1998,PLA2012QF} is switched in superconducting state, or at mechanical closing of superconducting loop \cite{PRB2001}. We can not doubt that the electric current (3) must appear in a ring or loop at $\Phi \neq n'\Phi_{0}$ when the wave function $\Psi = |\Psi |e^{i\varphi }$ describes real density of superconducting  pairs in all its segment. Quantum mechanics explains the numerous observations of the persistent current (3) with Bohr's quantization (1) and Aharonov-Bohm effect. But no theory can say how the mobile charge carriers can accelerate in direction opposite to the electromagnetic force. Superconducting  pairs (the mobile charge carriers) accelerate in accordance with the Newton's second law $mdv/dt = qE$ when the externally produced magnetic field increases in time $d\Phi _{ext}/dt = SdB/dt = (L_{f} + L _{k})d I_{p}/dt$: $d(\Phi _{ext}-L_{f}I_{p})/dt = El = L _{k}d I_{p} =(lm/q)dv/dt$. The electric current falls down to zero after switching of a ring segment in the normal state with a resistance $R$ also in accordance with the Newton's second law \cite{PLA2012QF}. But the persistent current $ I_{p} = (n\Phi_{0} - \Phi )/L _{k}$ appears contrary to the Newton's second law when the ring segment returns to superconducting state \cite{PLA2012QF}. 

This puzzle is consequence of the well-known difference between superconductivity (as macroscopic quantum phenomenon) and perfect conductivity. The Meissner effect discovered as far back as 1933 \cite{Meissner1933} is the first experimental evidence of this difference and this puzzle. Therefore the astonishment express by Jorge Hirsch is valid: "{\it Strangely, the question of what is the 'force' propelling the mobile charge carriers and the ions in the superconductor to move in direction opposite to the electromagnetic force in the Meissner effect was essentially never raised nor answered}" \cite{Hirsch2010}. 

Quantum description of macroscopic quantum phenomena is at least incomplete without such force. A. Einstein, B. Podolsky and N. Rosen \cite{EPR1935} have shown eighty years ago that quantum-mechanical description of physical reality is not complete assuming the impossibility of 'spooky action at a distance' \cite{Bell1981}.  The incompleteness considered here reveals that quantum mechanics predicts 'spooky action at a distance' of other type: the persistent current (3) should appear in a ring segment $l _{B}$ when a spatially separated segment $l _{A}$ is switched in superconducting state, see Fig.1 of \cite{PLA2012QF}. This action at a distance is spooky because of the impossibility to deduce from quantum mechanics a force which could accelerate superconducting  pairs in the segment $l _{B}$. In contrast to the case of the EPR correlation this action at a distance must not be instantaneous. On the other hand this action is real in contrast to the EPR correlation. 

The non-locality of the EPR correlation is deduced logically from Born's interpretation and Dirac's jump. The quantum state of Bob's particle spatially separated from Alice should change at her observation of her particle of the EPR pair (from (4) to (6) in \cite{Nikulov2013}) because, as Dirac postulated, "{\it a measurement always causes the system to jump into an eigenstate of the dynamical variable that is being measured}" \cite{Dirac1930}. The EPR correlation (i.e. "{\it entanglement of our knowledge}" according to Schrodinger \cite{Schrod35E} ) presupposes "{\it an unavoidable and uncontrollable impression from the side of the subject onto the object"} \cite{SchrodingerHum}. The wave function describes first of all the knowledge of the observer (Alice) according to Born's interpretation. Alice's knowledge (the subject) about both her and Bob's particles changes instantly due to her observation, from (4) to (6) in \cite{Nikulov2013}. The EPR correlation is possible due to Dirac's jump postulating the impression from the side of the subject (Alice's knowledge) onto the object (quantum state of the EPR pair). This correlation is non-local because our knowledge is non-local. The EPR correlation is spooky because "{\it it suggests some sort of psychokinetic effect of the conscious 'observer' on basic physical phenomena}" \cite{Deutsch1997FR} and {\it "The question cannot be ruled out as lying in the domain of psychology"} \cite{Everett1957}. 

In contrast to the EPR correlation neither the observer nor psychology are needed for the description of the quantization effects observed in superconductors. The realistic interpretation of the wave function proposed by Schrodinger rather than the subjective interpretation proposed by Born is used for description of superconductivity. The segment $l _{A}$ is switched in superconducting state because of a real physical influence, cooling for example, rather than of psychokinetic effect of the conscious observer. Dirac's jump is absent in this description. But we should postulate other jump until we can not say what is the force accelerating the mobile charge carriers against the electromagnetic force. 

The angular momentum of each electron pair changes on $\Delta m_{p} = \pm \hbar (n - \Phi /\Phi _{0})$ between  $m_{p} = q\Phi /2\pi $ and $\hbar n$ when the segment $l _{A}$, see Fig.1 of \cite{PLA2012QF}, is switched between superconducting and normal states. The total change $\Delta M _{p} = \pm N _{s}\Delta m_{p} = \pm (2m/q)I _{p}S$ is macroscopic because of the huge number of superconducting  pair $N _{s} = n _{s}s2\pi r \approx 10^{10}$ in a real ring. The electrical current $I(t) = I_{p}\exp -t/\tau_{RL}$ decays during a relaxation time $\tau_{RL} = L/R$ after the switching of the ring or ring segment $l _{A}$ in normal state with of a non-zero resistance $R > 0$. We know that the current decays and the angular momentum changes under the influence of the dissipation force $F _{dis} = -\eta v $. We can write the Newton's second law both for the segment $l _{A}$ switched in normal state $mdv/dt = qE + F _{dis} = RI(t)/l _{A} + F _{dis}$ and for the superconducting segment $l - l _{A}$: $mdv/dt = -qE  \approx  RI(t)/(l - l _{A}) $. The relaxation time $\tau_{RL} = L/R$ is deduced from these relations. But we can not deduce a time during which the electric current change from $I = 0$ to $I = I _{p}$ because quantum mechanics can not say what is the force changing the angular momentum of each electron pair from $m_{p} = q\Phi /2\pi $ to $\hbar n$. This time can be measured. Such experiment may be important. But no experimental result can eliminate the incompleteness of quantum mechanics. 

\subsection{Direct electric current can flaw against direct electric field.}
\label{sec:2.2}

\begin{figure}
\includegraphics{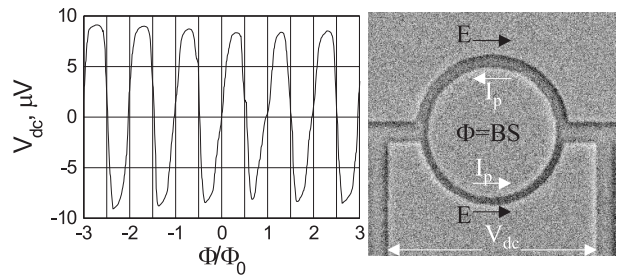}
\caption{\label{fig:epsart} The observations \cite{Letter07,toKulik2010,PerMob2001,Letter2003,PLA2012} of the quantum periodicity in the dc voltage give direct evidence of the paradox which can not be described completely: the persistent current $\overline{I _{p}}$ circulating clockwise or anticlockwise flows in one of the ring-halves against the dc electric field $E = -\nabla V_{dc}$ directed from left to right or from right to left. The photo of a real aluminium ring with the radius $r \approx  2 \ \mu m$ is shown on the right. Such ring was used for the observation of the $V _{dc}(\Phi )$ oscillations shown at the left.}
\end{figure}  

Quantum mechanics and the law of energy conservation predict a mechanical force acting between boundaries of Josephson junction interrupting superconducting loop \cite{PRB2001}. This mechanical force should depend periodically on magnetic flux inside the loop and can be measured \cite{PRB2001}. We can also deduce that the potential voltage with a direct component $V _{dc}$ can be observed on the segment $l _{A}$ when it is switched between superconducting and normal states with a frequency $f _{sw} = N _{sw}/\Theta $: $V _{dc} =  \int _{0}^{\Theta }dt V_{A}(t)/\Theta = \int _{0}^{\Theta }dt RI(t) =  \int _{0}^{\Theta }dt RI_{p}\exp {-t /\tau_{RL}} $ equals approximately $V _{dc} \approx  \sum^{i=N _{sw}}_{i=1}Lf _{sw} I_{p,i}/N _{sw} =  Lf _{sw} \overline{I_{p}}$ at a low frequency $f _{sw} \ll  1/\tau_{RL}$ and $V _{dc} \approx R \overline{I_{p}}$ at a high frequency $f _{sw} \gg  1/\tau_{RL}$ \cite{JLTP1998,PLA2012QF}. The dc voltage should oscillate with magnetic field likewise the average value of the persistent current $ \overline{I_{p}} = (\overline{n}\Phi_{0} - \Phi )/L _{k}$, where $\overline{n} = \sum _{n}nP _{n}(\Phi)$ and $P _{n}(\Phi)$ is the probability of the switching in superconducting state with the quantum number $n$ at magnetic flux inside the ring $\Phi $. Similar oscillations were observed at the switching of asymmetric ring between superconducting and normal states induced by the ac current \cite{PCJETP07,Letter2003} or a noise \cite{Letter07,toKulik2010,PerMob2001,PLA2012}. These observations give experimental evidence of a paradox. The dc electric current $\overline{I_{p}}$ flows against the dc electric field $E = -\nabla V _{dc}$ in one of the ring-halves, Fig.1.  

This paradox can be described if we take into account that the change of the angular momentum because of the dissipation force is equilibrated with the opposite change because of quantization (1), see \cite{PLA2012QF}. The change of momentum because of quantization reiterative many times $N _{sw}/\Theta $ in a time unity was called in \cite{PRB2001} 'quantum force'. The 'quantum force' was introduced in \cite{PRB2001} for the description of other paradox, observed both in the fluctuation region of superconducting rings \cite{Letter07,toKulik2010,LP1962} and normal metal rings \cite{PCScien09,PCPRL09}. It is well known that an electrical current induced in a resistive circuit will rapidly decay in the absence of an applied voltage. But the persistent current does not decay in resistive rings \cite{Letter07,toKulik2010,LP1962,PCScien09,PCPRL09}. The authors \cite{PCScien09,PCBirge2009} claim that this equilibrium current flowing through a resistive circuit is dissipationless. The author \cite{PCBirge2009} confesses that "{\it The idea that a normal, nonsuperconducting metal ring can sustain a persistent current - one that flows forever without dissipating energy - seems preposterous}". This idea is not only preposterous but also useless because it can not explain how the persistent current can flow against electric field, Fig.1. The both paradoxes can be described with the help of the 'quantum force' \cite{PRB2001}, i.e. taking into account the change of the angular momentum due to quantization \cite{PLA2012QF}. But this description cannot be considered complete because of the impossibility to answer on the question: "What is the 'force' propelling the mobile charge carriers to move in direction opposite to the electromagnetic force?"  

\section{Experimental results contradicting to the theoretical predictions }
\label{sec:3} 
The observations \cite{Letter07,toKulik2010,PCJETP07,PerMob2001,Letter2003,PLA2012} of the quantum oscillations of the dc voltage could be possible thanks to the opportunity to make asymmetric rings. Such opportunity is inconceivable in the case of atom. The peculiarities of the $V _{dc}(\Phi)$ oscillations can be described with the help of the relations (3) and (4) deduced from the quantization (1) and the Hamiltonian (2). The dc voltage $V _{dc}(\Phi)$ changes its sign at $\Phi = n'\Phi_{0}$ and $\Phi = (n'+0.5)\Phi_{0}$ because the average value of the persistent current $\overline{I_{p}} = \sum _{n}P _{n}(\Phi) (n\Phi_{0} - \Phi)/L _{k} = (\overline{n}(\Phi)\Phi_{0} - \Phi)/L _{k}$ should change its sign according to (3) and (4). The quantum state $n$ with the minimal energy (4) has the predominant probability $P _{n} \propto \exp{- E_{n}/ k _{B}T} $ because of the strongly discrete spectrum of permitted state of real superconducting rings: $|E_{n+1} - E_{n}| = |\Phi_{0}^{2}/ L _{k} + 2\Phi_{0}(n\Phi_{0}-\Phi)/ L _{k}| \gg k _{B}T$ at $\Phi \neq (n+0.5)\Phi_{0}$. The dc voltage $V _{dc}(\Phi)$ may be considered as the rectified ac voltage \cite{PCJETP07}. The rectification effect is observed due to the anisotropy of the critical current of asymmetric superconducting ring, i.e. the value of the external current $I _{ext}$ at which the ring is switched in the normal state depends on the $I _{ext}$ direction \cite{PCJETP07}. 

The ring is switched when the current density arrives at the critical value in one of the ring-halves. The persistent current $I _{p}$ (3) increases the current density in one of the ring-halves, Fig.1 in \cite{PCJETP07} and \cite{PRL06Rej}, and therefore decreases the critical value $I _{c}$ of the external current $I _{ext}$. The magnetic dependence of the critical value of the symmetric ring (with the same section of the ring-halves) $I_{c} = I_{c0} - 2|I_{p}| = I_{c0} - 2|n\Phi_{0} - \Phi |/L _{k}$, deduced from the quantization condition (1) and the Hamiltonian (2), predicts correctly the experimental results, see Fig.2 in \cite{JETP07J}. But measurements of asymmetric ring  (with the different section of the ring-halves, Fig.1) have discovered fundamental disagreements with the theoretical prediction, see the relation (2) and Fig.19 in \cite{PCJETP07} and Fig.2 in \cite{PRL06Rej}. According to the universally recognized point of view quantum periodicity is observed due to the change of the quantum number $n$ with magnetic flux at $\Phi = (n'+0.5)\Phi_{0}$. The pair velocity, see Fig.4.5 in \cite{Tinkham}, and the persistent current (3) should change linearly with $\Phi$ in the interval $(n-0.5)\Phi_{0} < \Phi < (n+0.5)\Phi_{0}$ from $ I_{p} = 0.5\Phi_{0}/L _{k}$ to $I_{p} = -0.5\Phi_{0}/L _{k}$ and should change by jump from $ I_{p} = -0.5\Phi_{0}/L _{k}$ to $I_{p} = 0.5\Phi_{0}/L _{k}$ at $\Phi = (n+0.5)\Phi_{0}$. Therefore the maximums of the critical current $ I_{c} (\Phi )$ should be observed at $\Phi = n\Phi_{0}$ and the minimums at $\Phi = (n+0.5)\Phi_{0}$. Measurements of the symmetric rings corroborate this prediction, see Fig.2 in \cite{JETP07J}. But the extreme values of the periodic dependence of the critical current $ I_{c+} (\Phi )$, $ I_{c-} (\Phi )$ of asymmetric rings measured in the opposite directions are shifted by a quarter of the flux quantum $ \Phi_{0}/4$ \cite{JETP07J}. The shift of $ I_{c+} (\Phi )$ and $ I_{c-} (\Phi )$ in the opposite direction provide the anisotropy of the critical current $I_{an} (\Phi ) = I_{c+} (\Phi ) - I_{c-} (\Phi ) \neq 0$ and explain the rectification effect \cite{PCJETP07}. But it contradicts the theoretical predictions according to which the extreme values can not be shifted and the jump of the critical current should be observed at $\Phi = (n+0.5)\Phi_{0}$.

The absence of the jump is a most fundamental contradiction between theory and experiment. According to the Bohr's quantization (1) the quantum number $n$ describing angular momentum must be integer. Therefore the change of this number must result in the jump of the persistent current (3) equal $\Phi_{0}/L _{k}$. This jump should not result in the jump of the critical current of the symmetric ring but it must result to the $I_{c}$ jump in the case of the asymmetric ring \cite{PCJETP07,PRL06Rej} and the ring with asymmetric link-up of current leads \cite{NANO2011}. The quantum periodicity in the critical current observed in \cite{PCJETP07,PRL06Rej,NANO2011} testifies to the change of the quantum number $n$. But the continuity of the $I_{c}$ dependence does not allow to say at which value of the magnetic flux the quantum number could be change on the unity. The jump of the critical current connected with the $n$ change was observed at measurements of more complicated structure \cite{JETP2011}. 

The contradictions were revealed also at measurements of the persistent current. According to (3) and (4) the two permitted states $n$ and $n+1$ with the non-zero persistent current $I _{p,n} \approx -0.5\Phi_{0}/L _{k}$ and $I _{p,n+1} \approx 0.5\Phi_{0}/L _{k}$ should be observed at $\Phi \approx (n+0.5)\Phi_{0}$. These two states were observed, for example, at measurements of the magnetization $\Delta \Phi _{Ip} = L\overline{I_{p}}$ of flux quantum bit (qubit), i.e. superconducting loop with three Josephson junctions, see Fig.4 in \cite{1Shot02}. But the observations of a $\chi $-shaped crossing of the  $I_{p,n}( \Phi)$ and $I_{p,n+1}( \Phi)$ dependencies, see Fig.4 in \cite{1Shot04}, reveal the contradiction with the theoretical prediction. According to the Bohr's quantization, states with the persistent current $-0.5\Phi_{0}/L _{k} < I _{p} < 0.5\Phi_{0}/L _{k}$ must be forbidden at $\Phi = (n+0.5)\Phi_{0}$. The authors \cite{1Shot04} interpret the $\chi $-shaped crossing as the single-shot readout of macroscopic quantum superposition of flux qubit states. But this claim as well as the experimental observations \cite{1Shot04} of the states with $I _{p} = 0$ forbidden at  $\Phi = (n+0.5)\Phi_{0}$ contradict to the orthodox quantum mechanics. 

\section{No theory can describe two opposite cases using the same Hamiltonian }
\label{sec:4}
The explanation of the quantum periodicity, considered above, is based on the assumption that the kinetic energy is the total energy of the persistent current which depends on magnetic field. But it is well known that electric current $I _{p}$ circulating in the ring with the area $S$ induces magnetic dipole moment equal $M _{m} = I _{p}S$ which has the energy equal  $E _{M} = - M _{m}B = I _{p}\Phi $ in an externally produced magnetic field $B$. The total energy $ E _{t} = E _{k} + E _{M}$ of the persistent current must be equal 
$$E _{t} = E _{k} + E _{M} =  \frac{n\Phi_{0}^{2} - \Phi ^{2}}{2L_{k}} \eqno{(5)}$$
According to (5), in contrast to (4), the diamagnetic state has minimal energy at any magnetic flux $\Phi = BS$, the quantum number $n$ should not change with $\Phi $ and the quantum periodicity should not be observed \cite{PhysicaC2015}. 

\subsection{We must challenge the conventional description of the quantum periodicity }
\label{sec:4.1}
The energy of the magnetic dipole moment was not taken into account in the theory of quantization \cite{QuTh1961} because only the kinetic energy of the current can be deduced from the canonical Hamiltonian (2) \cite{Leggett2014}. The energy $E _{M} = - M _{m}B = I _{p}\Phi $ can not be deduced from the Hamiltonian both in quantum and classical case \cite{Why2015}. But it is well known that this energy exists. It is enough easy to show in the classical case that the total energy of the $I _{p}$ state in an externally produced magnetic field $B$, defined as the energy expended for the creation of this state,  should be equal the sum $E _{t} = E _{k} + E _{M} + E _{f}$ of the kinetic energy $E _{k} = L _{k}I _{p}^{2}/2$, the energy $E _{M} = I _{p}\Phi $ of the magnetic dipole moment $M _{m} = I _{p}S$ in magnetic field $B$ and the energy $E _{f} = L _{f}I _{p}^{2}/2$ of magnetic field induced by the current $ I _{p}$ \cite{Why2015}. Since the energy due to the field term $E _{f} = L _{f}I _{p}^{2}/2$ is less than the kinetic energy of the current by a factor of the order of the ratio of the cross-sectional area of the conductor $s$ to $\lambda _{L}^{2}$, we can always neglect it for a sufficiently thin conductor \cite{Tinkham} p.123. This approximation of weak screening $L_{f}  \approx \mu _{0}l \ll L _{k} \approx (\lambda _{L}^{2}/s) \mu _{0}l$ is valid for the description of the quantum periodicity \cite{Letter07,toKulik2010,PCScien07,JETP07J,PCJETP07,PerMob2001,Letter2003,PLA2012} observed at measurements of sufficiently thin conductor with the cross-sectional area $s \ll \lambda _{L}^{2}$. The energy $E _{f} = L _{f}I _{p}^{2}/2$ is less than $E _{k} = L _{k}I _{p}^{2}/2$ but the energy of the magnetic dipole moment $E _{M} = I _{p}\Phi $ is not less at $L_{f} \ll L _{k}$. Our naive tendency to identify the Hamiltonian with the energy misleads \cite{Leggett2014}. 

The energy of the magnetic dipole moment $E _{M} = I _{p}\Phi $ is deduced from the history, "{\it involving time-dependent forces}" \cite{Leggett2014}, of the state rather than from the Hamiltonial  \cite{Why2015}. The momentum and the velocity of the mobile charge carriers change under influence of the known forces in the case of perfect conductivity \cite{Why2015}. Therefore the energy $E _{M} = I _{p}\Phi $ is easy deduced in the classical case \cite{Why2015}. Such deduction is not possible in the quantum case because of the incompleteness of quantum mechanics considered above. Quantum mechanics can not describe the history of the current state (3) involving time-dependent forces \cite{Why2015}. Nevertheless we can deduce the existence of the energy $E _{M} = I _{p}\Phi $ also in the quantum case using experimental data \cite{Why2015}. The persistent current $I_{p}$ of flux qubit \cite{1Shot02,1Shot04}, superconducting ring \cite{PCScien07} and normal metal ring \cite{PCPRL09} was measured with the help of the measurement of the additional magnetic flux $\Delta \Phi _{Ip} = L _{f}I_{p}$. Consequently a change of the persistent current $I_{p}$ in a ring with the magnetic inductance $ L _{f}$, should induce Faraday's voltage $-d\Phi _{Ip}/dt = - L _{f}dI_{p}/dt$ in the first loop creating magnetic flux, for example, $\Phi _{0}/2$, see Fig.2Qu in \cite{Why2015}. We can not say during which time the current $ I_{p}$ can change its direction. But the power source inducing the magnetic flux $\Phi _{0}/2$ should expend the additional energy $2I _{p}\Phi $ in any case \cite{Why2015}. Thus, we must conclude that the energy of the two permitted states of superconducting ring $n$ and $n+1$ should differ and the total energy should be described by the relation (5) rather than (4). If we can not doubt in the law of energy conservation. The requirement of this law challenges the conventional description of the quantum periodicity observed in numerous works \cite{Letter07,toKulik2010,PCScien07,JETP07J,PCJETP07,PerMob2001,Letter2003,PLA2012}. 
 
\subsection{We must challenge the description of the Zeeman effect}
\label{sec:4.1}
The energy of the magnetic dipole moment in magnetic field must be absent so as to quantum mechanics could describe the quantum periodicity considered above. But this energy must exist in order to atomic phenomena could be described. According to the predominate belief quantum mechanics describes successfully the both phenomena. But how could this description be possible if the magnetic dipole moment in magnetic field can not be deduced from the canonical Hamiltonian? How could Dirac explain the Zeeman effect in his book \cite{Dirac1930} published first as far back as 1930? Dirac used other definition of the operator of the canonical momentum and the Hamiltonian different from the one \cite{LandauL} prevalent now. 

Richard Feynman in the Section "The Schrodinger Equation in a Classical Context: A Seminar on Superconductivity" of his Lectures on Physics \cite{FeynmanL} writes about "Two kinds of momentum": "{\it It looks as though we have two suggestions for relations of velocity to momentum, because we would also think that momentum divided by mass, $\hat{p}/m$, should be a velocity. The two possibilities differ by the vector potential. It happens that these two possibilities were also discovered in classical physics, when it was found that momentum could be defined in two ways. One of them is called "kinematic momentum," but for absolute clarity I will in this lecture call it the "$mv$-momentum." This is the momentum obtained by multiplying mass by velocity. The other is a more mathematical, more abstract momentum, sometimes called the "dynamical momentum," which I'll call "$p$-momentum"$\cdot \cdot \cdot $ It turns out that in quantum mechanics with magnetic fields it is the p-momentum which is connected to the gradient operator $\hat{p}$, so it follows that (21.13) is the operator of a velocity}". The operator of a velocity according to the relation (21.13) of the Feynman Lectures \cite{FeynmanL} is $(\hat{p} - qA)/m$, where $\hat{p} = -i\hbar \nabla = -i\hbar (i_{x}\partial /\partial x +i_{y}\partial /\partial y + i_{z}\partial /\partial z )$ corresponds to the prevalent definition. 

But Dirac defined that the gradient operator $-i\hbar \nabla $ is the operator of the '$mv$-momentum' rather than  '$p$-momentum'. He writes in the beginning of the section 41. "The Zeeman effect for the hydrogen atom" of \cite{Dirac1930}:  "{\it We shall now consider the system of a hydrogen atom in a uniform magnetic field. The Hamiltonian (57) with $V = -e^{2}/r$, which describes the hydrogen atom in no external field, gets modified by the magnetic field, the modification, according to classical mechanics, consisting in the replacement of the components of momentum, $p _{x}$, $p _{y}$, $p _{z}$, by $p _{x}+qA _{x}$, $p _{y}+qA _{y}$, $p _{z}+qA _{z}$, where $A _{x}$, $A _{y}$, $A _{z}$ are the components of the vector potential describing the field}". The operator of the "$p$-momentum" is $\hat{P} = \hat{p} +qA = -i\hbar \nabla + qA$ according to Dirac's definition. According to the classical definition (see the relation (16.10) in \cite{LLFT}) and (2) the term $(\hat{P} - qA)^{2}/2m = \hat{p}^{2}/2m = (-i\hbar \nabla)^{2}/2m $ should be in the Hamiltonian. The energy of the magnetic dipole moment in magnetic field can not be deduced from such Hamiltonian. Dirac used other definition of the Hamiltonian: "{\it For a uniform field of magnitude $B$ in the direction of the z-axis we may take $A _{x} = - By/2$, $A _{y} = Bx/2$, $A _{z} = 0$. The classical Hamiltonian will then be}" \cite{Dirac1930}   
$$H = \frac{1}{2m}[(p_{x}  - \frac{1}{2}qBy)^{2} +(p_{y} + \frac{1}{2}qBx)^{2} + p_{z}^{2}] - \frac{q^{2}}{r}   \eqno{(6)}$$ 
Dirac could deduced the energy of the magnetic dipole moment of atom in magnetic field only due to this non-canonical definition of the momentum and the Hamiltonian (88): "{\it If the magnetic field is not too large, we can neglect terms involving $B^{2}$, so that the Hamiltonian (88) reduces to} \cite{Dirac1930}  
$$\hat{H} = \frac{1}{2m}(\hat{p_{x}}^{2} +\hat{p_{y}}^{2} + \hat{p_{z}}^{2}) - \frac{q^{2}}{r} +  \frac{qB}{2m}\hat{m} _{z}  \eqno{(7)}$$
The relation (7) corresponds to the relation (89) in \cite{Dirac1930} without the spin term $\hbar \sigma _{z}$. $\hat{m} _{z} = x\hat{p_{y}}-y\hat{p_{x}}$ is the operator of z-component of the orbital angular momentum of atom. The extra terms due to the magnetic field $(qB/2m)\hat{m} _{z}$ describes the energy of the magnetic moment $(q/2m) \Psi ^{*}\hat{m} _{z}\Psi $ in the magnetic field $B$ according to Dirac \cite{Dirac1930}. "{\it The external magnetic field splits the atomic levels and removes the degeneracy with respect to the directions of the total angular momentum (the Zeeman effect)}" \cite{LandauL}. 

\subsection{Quantum mechanics can not describe the both opposite cases}
\label{sec:4.2}
Strangely, the direct opposite of the phenomena observed at measurements of atoms and superconducting rings in magnetic field was never before noticed. The effect of splitting a spectral line of atoms into several components in the presence of a static magnetic field discovered by Pieter Zeeman as far back as 1896 testifies to the existence of the energy of the magnetic moment in the magnetic field. It is well known that this energy must be also in the case of the electric current circulating in the ring clockwise or anticlockwise. But the quantum periodicity in different parameters \cite{Letter07,toKulik2010,PCScien07,JETP07J,PCJETP07,PerMob2001,Letter2003,PLA2012} can not be described if this energy is taken into account. Most physicists believed during a long time that quantum mechanics describes successfully the both opposite cases. But we must admit that the both phenomena can not be described consistently. In order to describe the quantum periodicity in the persistent current (3) we must explain why the energy $E _{M} = - M _{m}B = I _{p}\Phi $ could not be taken into account and how the two permitted states $n$ and $n+1$ could have the same energy at $\Phi = (n+0.5)\Phi_{0}$ if the change of the persistent current (3) from $I _{p} = -0.5\Phi_{0}/L _{k}$ to $I _{p} = 0.5\Phi_{0}/L _{k}$ should induce Faraday's voltage $-d\Phi _{Ip}/dt = - L _{f}dI_{p}/dt$. 

The description of the Zeeman effect is doubtful because of the non-canonical definition used by Dirac  \cite{Dirac1930}. According to Dirac's definition $\hat{p} = m\hat{v} = -i\hbar \nabla $ the persistent current $I_{p} = (sq/m2\pi r)\oint_{l}dl \Psi ^{*}(-i\hbar \nabla)\Psi  = n\Phi_{0}/L_{k}$ should not depend on magnetic flux $\Phi$ inside the ring. The Aharonov - Bohm effect \cite{AB1959} and other known phenomena also should not be observed. Quantum mechanics seems to describe successfully different quantum phenomena due to the different definitions of the canonical momentum and the Hamiltonian. A consequence of these different definitions may be observed in the section XV. "Motion in a Magnetic Field" of the book \cite{LandauL}. The Hamiltonian (113.1) was written as in Dirac book $(\hat{p} + eA)^{2}/2m$ in the paragraph 113 "An atom in a magnetic field", whereas the canonical definition $(\hat{p} - eA)^{2}/2m$ was used in the relations (111.3), (111.4), (115.2) of all other paragraphs of this section. This inconsistency was in the Russian edition 1963 and was eliminated in the posterior editions of  \cite{LandauL}. Editors have written $(\hat{p} + |e|A)^{2}/2m$ instead of $(\hat{p} + eA)^{2}/2m$, where $-|e| =e$ is the electron charge. It is obvious that $(\hat{p} + |e|A)^{2}/2m \equiv (\hat{p} - eA)^{2}/2m $. But the editors "correcting typo" did not notice that $(\hat{p} - eA)^{2}/2m $ is only the kinetic energy and the energy of the orbital angular momentum of the atom can not be deduced from this Hamiltonian without a mathematical mistake. 

According to the elementary mathematics the equality $(\hat{P} - eA)^{2}/2m = m\hat{v}^{2}/2$ must be deduced from the equality $\hat{P} - eA = m\hat{v}$. The additional summand $\mu _{B}\hat{L}B$ could appear in the relation (113.2) of the book \cite{LandauL} due to the illegal substitution of $\hat{P}^{2}/2m$ by $m\hat{v}^{2}/2$ in $\hat{H} _{0}$. Here $\mu _{B}$ is the Bohr magneton; $\hbar \hat{L}$ is the operator of the total orbital angular momentum of the atom. Dirac did not use this illegal substitution due to the non-canonical definition of the momentum and the Hamiltonian used in \cite{Dirac1930}. He defines the Hamiltonian and the kinetic energy in the different ways: $\hat{P}^{2}/2m = (\hat{p} +qA)^{2}/2m = (m\hat{v} +qA)^{2}/2m $ is written in the Hamiltonian (88) whereas the kinetic energy is written as $\hat{p}^{2}/2m = m\hat{v}^{2}/2$ in the relation (89) of the book \cite{Dirac1930}. Neither Dirac nor anybody could deduce the energy of the orbital angular momentum $\mu _{B}\hat{L}B$ using the canonical definition of the Hamiltonian and the kinetic energy.  Only the kinetic energy can be deduced from the Hamiltonian according to the canonical definition \cite{Why2015}. We must admit that the explanation of the Zeeman effect by Dirac \cite{Dirac1930} is doubtful because of groundlessness and inconsistence of his non-canonical definition of the momentum and the Hamiltonian. Quantum electrodynamics should also be challenged because Dirac used the same non-canonical definition in his relativistic theory of the electron \cite{Dirac1930}.

\section{Conclusion}
\label{sec:5}
The history of the orthodox quantum mechanics has begun with a problem. The wave theory proposed by Schrodinger has allowed to describe successfully atomic phenomena. Schrodinger tried to replace particles by wavepackets. But wavepackets diffuse. Schrodinger interpreted his wave function as a real wave \cite{Schrodinger1926} and defended this realistic interpretation \cite{Schrodinger1952}. But we can not think that a real density $|\Psi(r) |^{2}$ can change because of our observation. On the other hand we know from our everyday experience that the uncertainty of the next observation decreases after the first observation. Our experience convinces us that we will see a thing approximately in the same place $r$ at the second observation in which we saw it at the first observation. Therefore we are fully confident that the probability of observation changes from $|\Psi(r)|^{2} < 1$ to $|\Psi(r) |^{2} = 1$ after the first observation. Just therefore most physicists have rejected the Schrodinger's interpretation and have accepted Born's interpretation. 

But they did not take into account that the probability of observation $|\Psi(r) |^{2}$ changes in our mind according to our experience. The knowledge of the observer about the object changes at the observation. But the problem with wavepackets can not be solved if only the knowledge changes. Therefore Dirac jump \cite{Dirac1930}, wave function collapse \cite{Neumann1932}, or {\it "'quantum jump'… from the 'possible' to the 'actual'"} \cite{Heisenberg1959} was postulated. Dirac jump represents "{\it an unavoidable and uncontrollable impression from the side of the subject onto the object}" \cite{SchrodingerHum}. The non-locality of the EPR correlation \cite{EPR1935}, violation of Bell's inequality \cite{Bell1981}, the problem of free will \cite{Conway2006,Hooft2007} and other fundamental obscurities in quantum mechanics may be deduced logically from this postulation of the subjectivity according to which the change of our knowledge can change instantly the state of a distant quantum system. 

But most physicists refused to confess that the question of observation {\it "cannot be ruled out as lying in the domain of psychology"} \cite{Everett1957}. They, as well as the author \cite{Kampen1988}, "{\it dismissed out of hand the notion of von Neumann, Pauli, Wigner - that 'measurement' might be complete only in the mind of the observer}" \cite{Bell1989}. This mass delusion could be possible because of logical inconsistency of some statements postulated by the creators of quantum mechanics. According to the quantum postulate proposed by Bohr {\it any observation of atomic phenomena should include an interaction they with equipment used for the observation which can not be neglected} \cite{Bohr1928}. Bohr and Heisenberg (the uncertainty microscope \cite{Heisenberg1927}) substantiated the impossibility to measure simultaneously some variables with any great degree of accuracy by the indeterminacy introduced by this interaction with equipment. 

Dirac, on the one hand, following to Bohr and Heisenberg, stated that "{\it it is not in general permissible to consider that two observations can be made exactly simultaneously, and if they are made in quick succession the first will usually disturb the state of the system and introduce an indeterminacy that will affect the second}" \cite{Dirac1930}. But on the other hand he should postulate that the indeterminacy in quick succession of two observations of the same dynamical variable must disappear: "{\it When we measure a real dynamical variable $\xi $ the disturbance involved in the act of measurement causes a jump in the state of the dynamical system. From physical continuity, if we make a second measurement of the same dynamical variable $\xi $ immediately after the first, the result of the second measurement must be the same as that of the first. Thus after the first measurement has been made, there is no indeterminacy in the result of the second. Hence, after the first measurement has been made, the system is in an eigenstate of the dynamical variable $\xi $, the eigenvalue it belongs to being equal to the result of the first measurement. In this way we see that a measurement always causes the system to jump into an eigenstate of the dynamical variable that is being measured, the eigenvalue this eigenstate belongs to being equal to the result of the measurement}" \cite{Dirac1930}. 

Dirac's statement that "{\it after the first measurement has been made, there is no indeterminacy in the result of the second}" contradicts to the belief predominant up to now and voiced, in particular, by the author \cite{Kampen1988} that "{\it the quantum mechanical measurement is terminated when the outcome has been macroscopically recorded}" and that "{\it the mind of the observer is irrelevant}", see  \cite{Bell1989}. We can think that an interaction with equipment will introduce an indeterminacy. Bohr and Heisenberg postulated just this statement. But we can not think that a physical interaction will eliminate the indeterminacy. The idea of Dirac jump originates in our everyday experience which has convinced us that there is no indeterminacy in the result of the second observation. Heisenberg's statement {\it "Since through the observation our knowledge of the system has changed discontinuously, its mathematical representation also has undergone the discontinuous change and we speak of a 'quantum jump'"} \cite{Heisenberg1959} is also based on our experience. But according to our experience such 'quantum jump' describes a process of psychology rather than physics. Dirac has entangled psychology and physics postulating that the jump in our knowledge of the system excites the jump in the state of the system. 

The entanglement of physics and psychology is deduced logically from Born's interpretation. Therefore it is important to draw reader's attention that most quantum phenomena are described with the help of Schrodinger's interpretation \cite{Nikulov2013}. It is important now because of the efforts of some contemporary authors to solve the problems of quantum mechanics deduced from Born's interpretation, such as EPR correlation, violation of Bell's inequality and others. These efforts will be useless without consciousness of the fundamentally different problems of quantum mechanics considered here. The impossibility to describe the quantum periodicity and the Zeeman effect with the help of the same Hamiltonian has no relation to Born's interpretation. The self-contradictions of the orthodox quantum mechanics are various. They must be uncovered and appreciate in order to we have a chance to create a consistent and complete theory of quantum phenomena.


\begin{thebibliography}{99}
\bibitem{Leggett2014} I could understand that the energy of the magnetic dipole moment can not be deduced from the canonical Hamiltonian thanks to my private correspondence with Prof. Anthony Leggett who has surmised soundly that "{\it Perhaps our naive tendency to identify the Hamiltonian with the 'energy' is (as in some cases involving time-dependent forces) misleading?}"
\end{thebibliography}
\end{document}